# A centralized reinforcement learning method for multi-agent job scheduling in Grid

Milad Moradi
Department of Electrical and Computer Engineering
Isfahan University of Technology
Isfahan 84156-83111, Iran
mmoradi.research@gmail.com

*Abstract:*

One of the main challenges in Grid systems is designing an adaptive, scalable, and model-independent method for job scheduling to achieve a desirable degree of load balancing and system efficiency. Centralized job scheduling methods have some drawbacks, such as single point of failure and lack of scalability. Moreover, decentralized methods require a coordination mechanism with limited communications. In this paper, we propose a multi-agent approach to job scheduling in Grid, named Centralized Learning Distributed Scheduling (CLDS), by utilizing the reinforcement learning framework. The CLDS is a model free approach that uses the information of jobs and their completion time to estimate the efficiency of resources. In this method, there are a learner agent and several scheduler agents that perform the task of learning and job scheduling with the use of a coordination strategy that maintains the communication cost at a limited level. We evaluated the efficiency of the CLDS method by designing and performing a set of experiments on a simulated Grid system under different system scales and loads. The results show that the CLDS can effectively balance the load of system even in large scale and heavy loaded Grids, while maintains its adaptive performance and scalability.

## I. Introduction

One of the most significant problems in computer sciences, which is also important in economics, is multi-agent resource allocation. It addresses the problem of distributing a number of items among a number of agents [1]. The resource allocation, also called scheduling, is relevant to a variety of applications, such as Grid computing, public transportation, and network routing.

In recent decades, with the increasing demand for distributed computing, Grid computing emerges as a leading technology for supporting complicated and decentralized computing problems. Grid computing delivers a set of capabilities including aggregation, selection and sharing a number of heterogeneous resources that are distributed among geographically diverse locations [2]. One of the crucial tasks that directly affects the performance of Grid systems is coordinated resource allocation and job scheduling. By employing an efficient and robust algorithm for job scheduling, a Grid system can deliver its peak performance. When an applicable policy is used for resource allocation, Grid system achieves the speedup in job processing and provides high-quality services to users [3]. One of the key issues in Grid systems is load balancing that is defined as completing all the jobs at hand as soon as possible [2]. An adaptive scheduling method, which balances the load of Grid system effectively among the heterogeneous resources, is a requirement to maintain the performance of Grid system at a desirable level.

In recent years, many advances have been done in Grid job scheduling algorithms. These algorithms are divided into centralized and decentralized job scheduling. As centralized scheduling systems, which also known as traditional systems, we can point to PBS [4], SGE [5] and Condor [6]. These systems work effectively by utilizing global state information that is obtained from Grid environment. However, centralized resource allocation has a major drawback, i.e. single point of failure, which is also referred to as the lack of fault-tolerance ability. Furthermore, these methods is not scalable. To overcome the mentioned problems, researchers turning their approach to decentralized scheduling algorithms. However, other challenges were appeared by introducing the decentralized job scheduling methods. Most of decentralized schedulers, like AppleS [7] or Condor-G [8], may encounter many synchronization problems in resource management. These problems arise due to the fact that scheduling policies are performed individually by each scheduler, regardless of the other schedulers' decisions. In order to tackle these problems, some scheduling methods were proposed based on coordination mechanisms, like Condor Flock P2P [9]. However, a deficiency that brings a high communication overhead to these schedulers is the extra dependency on negotiation among the schedulers. Hence, coordinating the scheduling among the decentralized schedulers, with a reduced communication cost, is an important driving force for new researches.

In recent years, a number of reinforcement learning-driven methods have been proposed for job scheduling in Grid environment. Some of these methods are designed for decentralized scheduling. It means that multiple agents perform the task of learning based on information obtained from previous actions, i.e. selection of resources and receiving feedback from Grid environment. The advantages of these methods are their low communication needs among the scheduler agents, and their scalability. Even though, the reinforcement learning is a proper framework for multi-agent job scheduling, it seems that multiple learner agents may capture the state of entire system and resources with some differences. However, multi-agent scheduling is regarded as an appropriate solution to overcome the single point of failure problem. Therefore, utilizing a centralized and effective learning approach in combination with distributed job scheduling, can resolve the single point of failure, and also improves the load balancing of Grid and the scalability of scheduling method.

In this paper, we propose a Grid job scheduling method, named Centralized Learning Distributed Scheduling (CLDS). In the CLDS method, multiple agents perform the task of scheduling and produce rewards for learning task according to scheduled jobs' information. In this method, only one agent is responsible for the task of learning. This agent shares its information of Grid with scheduler agents by a utility table, and gets the reinforcement signals from them to make a unified view of Grid system. The advantages of this strategy are: 1) the communication cost is reduced by a simple coordination strategy, 2) since there are multiple schedulers that perform the scheduling task, and each of which can be employed as learner agent, there is no single point of failure, and 3) since there is a unified learner agent, the scheduler agents only see a single and real view of the Grid system. We designed and accomplished a set of experiments on a simulated Grid system to show the effectiveness of the CLDS method, in terms of load balancing, compared with other scheduling methods.

The rest of this paper is organized as follows. Section II gives an overview of previous job scheduling approaches that use reinforcement learning and other learning frameworks. In Section III, we describe the CLDS job scheduling method. Section IV explains the evaluation methodology and discusses the evaluation results. Finally, we draw conclusion and describe future work in Section V.

## II. RELATED WORK

In Grid environment, the resources are heterogeneous, the performance of resources is vary and the applications are diverse. Therefore, an adaptive scheduling method is needed. Reinforcement learning is a machine learning method that is widely used to solve uncertain decision making problems. It obtains sub-optimal or near-optimal policies by interacting with the environment [10]. The reinforcement learning can solve the difficulties of Grid resource allocation problem by providing a model-free framework.

Tesauro et al. [11] propose a hybrid scheduling approach that benefits from the advantages of both queuing and reinforcement learning models, in open-loop and closed-loop traffics. In their method, a queuing model policy controls the system, while reinforcement learning trains offline on data collected from the system. They use reinforcement learning to train nonlinear function approximators instead of lookup tables, which enables scaling to larger state spaces. In [12], a general framework is proposed to perform dynamic resource allocation among multiple entities. It uses reinforcement learning in combination with fuzzy rule bases, and can be employed in an environment with and without any existing resource allocation policies. Zhang et al. [13] present a multi-agent learning algorithm for optimizing online resource allocation in cluster networks. Each agent decides according to two connected learning problems. The first problem is local allocation that refers to deciding which tasks to be allocated locally. The second problem is task routing that refers to deciding where to forward a task. A number of heuristic strategies is developed, in order to speed up the learning stage and to avoid weak initial policies.

A hybrid resource management method is proposed in [14] to improve the system reliability in Grid and Cloud computing. It employs a reinforcement learning method in combination with neural network to help the scheduler to deal with dynamicity in execution environment. In a previous work [15], a meta-scheduler is presented for job scheduling in computational Grids. It utilizes a fuzzy rule-based system to develop a Grid scheduling middleware. It incorporates a swarm intelligence method aimed at knowledge acquisition, in order to improve the ability of adapting to changes in the resources and applications conditions.

Wu et al. [2] propose a distributed learning algorithm called Ordinal Sharing Learning (OSL) based on the reinforcement learning framework. In the OSL method, each scheduler has a utility table and updates it in two steps. Firstly, it updates the table using the local rewards produced for the resources. Secondly, it updates the table using the utility table of its adjacent scheduler. After updating the utility table, the scheduler sends its table to the neighbor scheduler, and the neighbor agent performs updating and sending of the utility table likewise. Compared to the OSL, our proposed method employs a centralized learning strategy. In the CLDS method, there is a single learner agent and a single utility table. The scheduler agents send their produced rewards to the learner, and it updates the utility table using all the rewards. For the subsequent resource allocation decisions, the scheduler agents will use the updated utility table. With the use of this strategy, all the schedulers see the same view of Grid system.

As noted earlier, the centralized job scheduling methods, which use a single scheduler agent, may fail due to the single point of failure problem. Therefore, the fault tolerance ability of Grid system is violated. Moreover, in decentralized scheduling methods, an efficient and low cost coordination strategy is needed, in order to all schedulers decide according to a unified view of Grid system and the state of resources. We propose a multi-agent scheduling method that tackles the single point of failure problem. Moreover, a coordination strategy is adopted based on a centralized reinforcement learning approach and limited communications.

### III. THE PROPOSED METHOD

As mentioned earlier, in model-based job scheduling methods that mostly rely on the Grid Information System (GIS), schedulers may have inaccurate and time-delayed information about resources. To tackle this problem, an adaptive scheduling algorithm is deserved. Such an algorithm should not be dependent on an accurate model. To deal with this type of scheduling problem, we propose a centralized reinforcement learning method in combination with coordinated multi-agent job scheduling. An accurate view of the state of resources is available for all schedulers, because one learner agent performs the learning task and provides the information about Grid system to all the scheduler agents. Moreover, all the scheduler agents can undertake the learning task. As a result, our proposed method can also deal with the single point of failure problem in Grid system. In the following, after describing a general model for job scheduling in Grid, our proposed method is explained in detail.

Before introducing our proposed method, we describe a general job scheduling model in Grid, which has been widely used in the literature to assess job scheduling algorithms [16]. In this model, there are several users, resources, and schedulers. The users produce jobs and submit them to the schedulers. Different schedulers receive the jobs from users and allocate them to the resources in parallel. Each scheduler can allocate jobs to any of the resources. For the simplicity reasons, it is assumed that all the resources are computing resources. In general, the problem of decentralized job scheduling in Grids, can be modeled using a multi-agent job scheduling system [16]. This model is denoted as a 6-tuple *(A, R, P, S, C, JSR)*, where $A=\{a_1, …, a_N\}$ is a set of agents, $R=\{r_1, …, r_M\}$ is a set of resources, $P:A×N→[0, 1]$ is a job submission function, $S:A×N→\mathcal{R}$ is a probabilistic job size function, $C:A×N→\mathcal{R}$ is a probabilistic capacity function, and *JSR* is a job scheduling rule [2].

Although the above model is based on some abstractions, it still preserves the dynamicity, heterogeneity, and randomness features of Grid environment.

#### A. Centralized learning method

One of the reinforcement learning strategies for multi-agent job scheduling is based on multiple independent learners [2]. In such type of learning, there is no explicit communication among the agents, and each agent learns independently based on local state and local reward. However, this strategy may lead to an anomalous resource allocation, because there is no communication among agents and they do not have a real view of Grid system. In this case, the agents learn according to their local information, and a coordination mechanism seems to be required.

To tackle the above problem, in our proposed method, the scheduler agents submit their local rewards to a learner agent. In each time step, the learner agent collects the rewards and updates a utility table that holds the efficiency of selecting the resources. Then, the learner agent sends the updated utility table to the scheduler agents, and the schedulers make their resource allocation decisions according to the utility table. In the following, each step of the learning method is explained thoroughly.

*1) Generating local rewards*

In each time step, the agent $a_i$ receives job scheduling tasks from one or more users. It makes a resource allocation decision for each job based on the utility table. The resource allocation strategy will be explained in the next subsection. When an agent submits a job to a resource, it records an entry for the submitted job in the *Scheduled Job List*. In the entry, *Job ID*, *Job Size* and *Resource ID* are recorded, also additional information is recorded, such as *Starting Time* and *Completion Time*. The agent $a_i$ searches the *Scheduled Job List* to attain the completion information of the jobs and generate reward for each one of the resources. If a job is completed, a positive reward is generated for the corresponding resource and its record is removed from the list. For each finished job $j_k$ in the *Scheduled Job List*, a positive reward is produced for the corresponding resource $r_q$, based on the *Job Size*, *Starting Time* and *Completion Time*, as follows:

$$reward(r_q) = Job\ Size(j_k) / Time\ to\ Completion(j_k) \quad (1)$$

where $reward(r_q)$ is the produced reward for the resource $r_q$, $Job\ Size(j_k)$ is the size of job $j_k$ submitted to the resource $r_q$, and $Time\ to\ Completion(j_k)$ is the total time that the resource $r_q$ spent to process the job $j_k$ and is calculated as $Completion\ Time(j_k) - Starting\ Time(j_k)$. Each finished job produces a positive reward for the corresponding resource in proportion to its size and the total time between its starting and completion. In this way, for two jobs with equal size, the job which was finished in less time, produces a greater positive reward for the corresponding resource.

For each unfinished job $j_k$ in the *Scheduled Job List*, a negative reward (i.e. a penalty) is produced for the corresponding resource $r_q$, based on the *Job Size* as follows:

$$reward(r_q) = -1 / Job\ Size(j_k) \quad (2)$$

where $reward(r_q)$ is the produced reward for the resource $r_q$, and $Job\ Size(j_k)$ is the size of job $j_k$ submitted to the resource $r_q$. Each unfinished job produces a negative reward for the corresponding resource in proportion to its inversed size. In fact, an unfinished job with greater size produces a lower negative reward, because the bigger job needs more time to be completed.

If an agent has more than one job submitted to same resource in its *Scheduled Job List*, it calculates the sum of all produced positive and negative rewards for that resource and generates a single reward. After generating local rewards, each agent $a_i$ puts the rewards into a *Reward Vector* and submits it to the learner agent. The $q_{th}$ element in the *Reward Vector* holds the produced reward for the resource $r_q$.

*2) Updating the utility table*

In each time step, the learner agent receives the reward vectors from all scheduler agents, and updates the utility table *U*. Then, it sends the new utility table to the scheduler agents. The utility table *U* is a vector that its size is equal to the number of resources, where *U(q)* holds the efficiency of the $q_{th}$ resource. The learner agent updates the utility table *U* using all the reward vectors, as follows:

$$U(q) = (1 - \alpha) \times U(q) + \alpha \times \sum_i Reward\ Vector_i(q) \quad (3)$$

where *α* is learning factor, and *Reward Vector$_i$(q)* is the $q_{th}$ element, which holds the reward for the $q_{th}$ resource, in the reward vector generated by the $i_{th}$ agent. In (3), all the produced rewards for a resource are added together and used for updating the efficiency of the resource.

After updating the utility table, the learner agent sends it to the scheduler agents. At the next time step, the scheduler agents will use the updated utility table to allocate the resources. They will generate the rewards and submit the *Reward Vector* to the learner agent again. In the following subsection, we will explain how the scheduler agents do the job scheduling task.

*B. Multi-agent job scheduling*

At each time step, in addition to generating the rewards and receiving the updated utility table, each scheduler agent submits the jobs in its *Job Queue* to appropriate resources. This decision is made according to the utility table. Each agent $a_i$, for each job $J_m$ in its *Job Queue*, selects the resource which has the greatest utility value in the utility table and submits the job $J_m$ to it. As mentioned earlier, after submitting a job to a resource, the agent $a_i$ records an entry for the submitted job in the *Scheduled Job List*. Then, it removes the submitted job from its *Job Queue*.

In the proposed job scheduling method, there is only one learner agent that collects the reward vectors, updates the utility table, and sends the new utility table to the scheduler agents. However, there is no single point of failure in this method, because every scheduler agent can be employed as the learner agent. The learner agent has not any special capabilities, and can be replaced by each one of the scheduler agents.

As discussed above, the communication cost grows linearly with the number of agents, and it is still better than other methods that use traditional coordination mechanisms and need the communications with exponential order.

IV. EVALUATION AND RESULTS

We conducted a set of experiments to evaluate the performance of our CLDS job scheduling method on a simulated Grid system. We compared our CLDS method with other three job scheduling methods, i.e. Least Load Selection (LLS), Random Selection (RS), and Decentralized Min-Min Selection (DMMS) [17]. In the LLS method, an agent selects the resource with the least load and submits the current job in the queue to it. If multiple resources have the same minimum load, the agent selects one of them randomly. In the RS approach, an agent selects the resources and allocates them to the jobs in queue based on a uniform probability distribution. In the DMMS approach, each agent performs the scheduling task independently based on decentralized Min-Min algorithm, which is a benchmark scheduling algorithm for performance evaluation.

There is a metric, named Average Load of Resources (ALoR) [2], for evaluating the performance of Grid job scheduling methods. The ALoR can be properly used to assess the efficiency of job scheduling in the Grid model explained in Section III. The simulated Grid system which was used for the evaluations, was thoroughly based on the early discussed model. In the model, resources are different in their processing capacity *C*. For each resource, the processing capacity is determined as the inverse of CPU time required to perform a job of a unit length [2]. The capacity of resources is generated randomly from a uniform distribution in a given interval. Each resource has a queue for arriving jobs, and performs only one single job at a given time, based on First In First Out (FIFO) order. Moreover, each job has a length that is chosen randomly from a uniform distribution in a given interval. Considering the above assumptions, the ALoR is defined as follows [2]:

$$ALoR = (1 / |R|) \sum_r LoR_r = (1 / |R|) \sum_r (l\_total_r / C_r) \quad (4)$$

where *|R|* is the total number of resources, $LoR_r$ is the load of $r_{th}$ resource, $l\_total_r$ is the total length of jobs in the queue of $r_{th}$ resource, and $C_r$ is the capacity of $r_{th}$ resource.

In performance evaluation of job scheduling algorithms, a minimum value of ALoR is preferred. The lower ALoR refers to the better load balancing and system efficiency. We evaluated the performance of our proposed job scheduling method under three different system scales and two different system loads. The system scale is determined by the number of schedulers and resources, and the system load is determined by the proportion of the total length of jobs in schedulers' queue to the total capacity of resources. In Fig. 1 to Fig. 6, the results of performance evaluation are represented. Our proposed method is indicated as CLDS. The different system scales and loads, used for performance evaluation, are given below:

- Small scale (50 schedulers, and 200 resources), and medium load (60%). The results are represented in Fig. 1.
- Medium scale (150 schedulers, and 400 resources), and medium load (60%). The results are represented in Fig. 2.
- Large scale (300 schedulers, and 1200 resources), and medium load (60%). The results are represented in Fig. 3.
- Small scale (50 schedulers, and 200 resources), and heavy load (90%). The results are represented in Fig. 4.
- Medium scale (150 schedulers, and 400 resources), and heavy load (90%). The results are represented in Fig. 5.
- Large scale (300 schedulers, and 1200 resources), and heavy load (90%). The results are represented in Fig. 6.

As can be seen in Fig. 1 to Fig. 6, our proposed job scheduling method, i.e. the CLDS, and the DMMS perform well in load balancing task under different scales and loads. The RS method fails to carry out the job scheduling task, and the LLS method shows the best efficiency among the four methods.

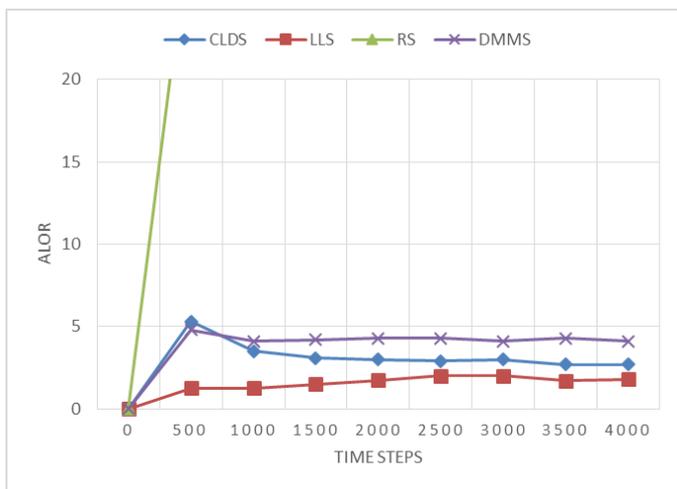

Fig. 1. ALoR for our proposed method (CLDS) and the other three methods, under a small scale and medium load system.

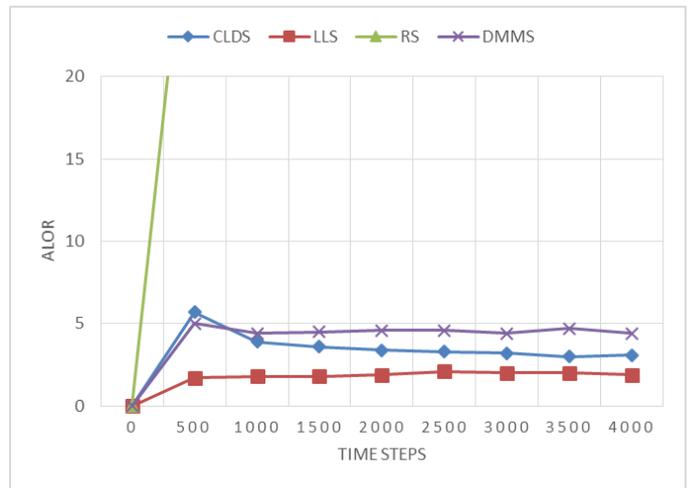

Fig. 2. ALoR for our proposed method (CLDS) and the other three methods, under a medium scale and medium load system.

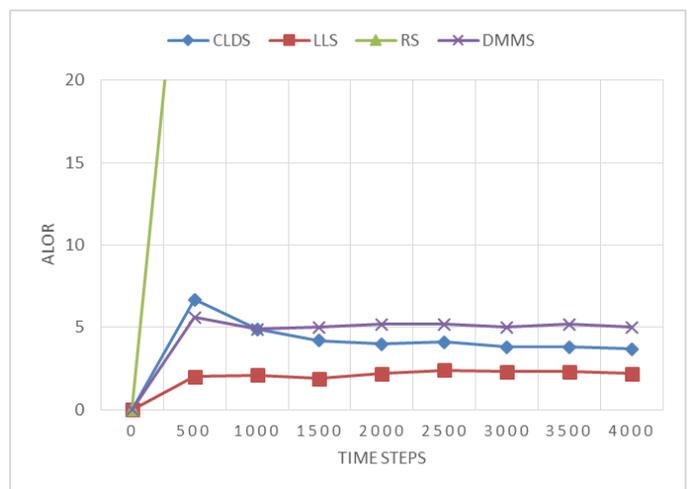

Fig. 3. ALoR for our proposed method (CLDS) and the other three methods, under a large scale and medium load system.

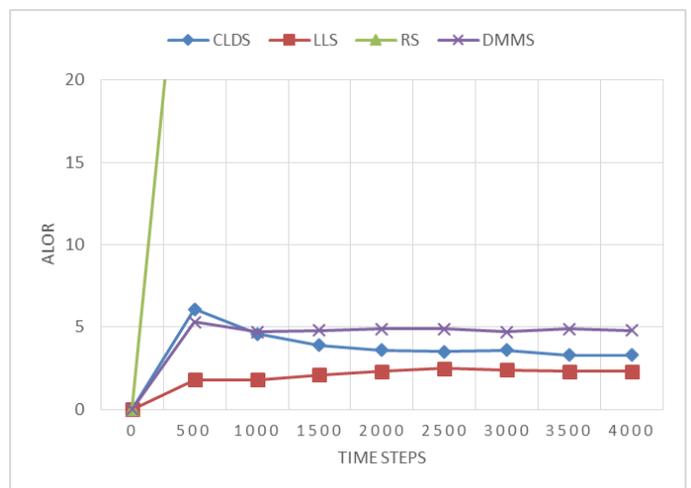

Fig. 4. ALoR for our proposed method (CLDS) and the other three methods, under a small scale and heavy load system.

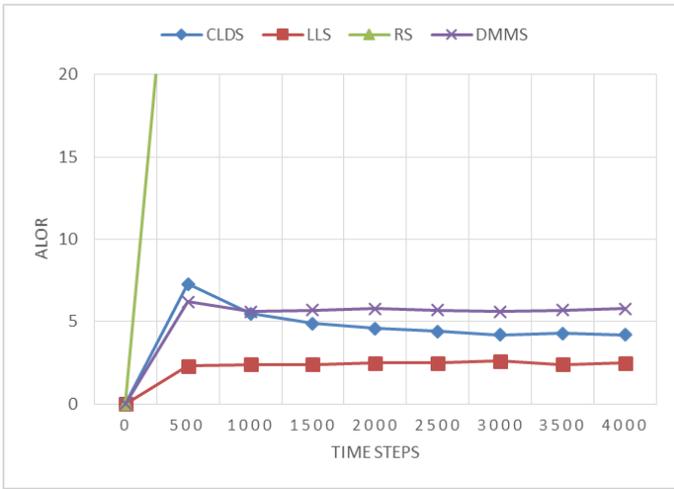

Fig. 5. ALoR for our proposed method (CLDS) and the other three methods, under a medium scale and heavy load system.

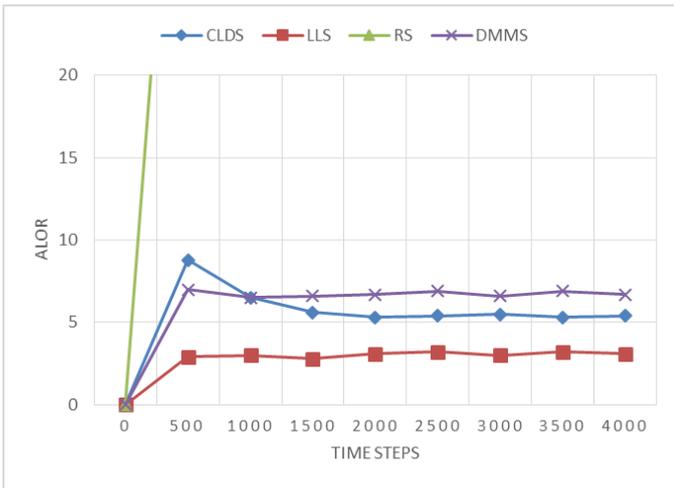

Fig. 6. ALoR for our proposed method (CLDS) and the other three methods, under a large scale and heavy load system.

At initial steps, our method performs worse than DMMS and LLS methods. It happens because in our method, the scheduler agents have no initial knowledge about the resources and the Grid system. At initial steps, the scheduler agents collect the information about the efficiency of resources, according to generated rewards and updating the utility table by the learner agent. After almost 1000 time steps, the CLDS begins to perform more efficiently than the DMMS. Afterwards, it converges to an efficient ALoR and obtains a sub-optimal policy for resource allocation. The LLS method achieves an impressive load balancing performance, however, it cannot be effectively employed in real world Grids. The LLS method has a high communication and computing cost. Moreover, it is a centralized scheduling method that is taken into account as a single point of failure. The DMMS method achieves the load balancing, although, the performance is lower than the CLDS and the LLS. In DMMS method, schedulers work without any coordination mechanisms, and therefore, some resources may not be utilized effectively or may be overloaded. The RS method does not consider the performance of resources and allocates the resources according to a random manner. Consequently, resources are not utilized based on their efficiency, the ALoR increases with a great degree, and the RS method fails to balance the load of resources.

We evaluated the effectiveness of our proposed job scheduling method under different system scales and system loads, in order to test its scalability and adaptive performance. As can be observed from the results, the CLDS method can converge to a sub-optimal policy in different system scales. The results show that the CLDS method is scalable and can be employed effectively in Grid systems with both small and large number of schedulers and resources. The CLDS method also provide an adaptive performance under medium and heavy system loads. When the load of system increases, the CLDS method still converges to a sub-optimal or even near-optimal policy.

## V. CONCLUSION

In Grid systems, a job scheduling method that can fairly balance the load of system among all resources is a main requirement to improve the performance of the entire system. Moreover, such scheduling method should have essential abilities, such as scalability, adaptive performance, and fault tolerance. In this paper, we proposed a multi-agent job scheduling method, named CLDS, for Grid system, using reinforcement learning framework. In the CLDS method, a single learner agent is responsible for receiving the produced rewards from multiple scheduler agents, updating a utility table that contains the efficiency of the resources, and sends the updated utility table to all the scheduler agents. The scheduler agents make resource allocation decisions according to the utility table. Each scheduler agent can be employed as learner agent, therefore, there would be no single point of failure in the Grid system. The results of experiments show that the CLDS method performs well in load balancing and converge to a sub-optimal policy, or even to a near-optimal policy in some cases. The coordination mechanism among the schedulers is established through a centralized reinforcement learning approach with limited communications, which makes a unified view of the state of resources to the schedulers. Therefore, the scheduler agents can allocate the resources more effectively. According to the evaluation results under different system scales and loads, the CLDS method shows an adaptive and scalable performance.

In future work, we will try to improve the CLDS method by incorporating a hierarchical scheduling mechanism.